\documentclass{aa}
\usepackage{subcaption}
\usepackage{sidecap}
\usepackage{txfonts}
\usepackage{natbib}
\bibpunct{(}{)}{;}{a}{}{,} 

\title{Stellar chemistry and planet size: insights from GALAH~DR4}
\author{ 
N.\ Sussholz\inst{\ref{inst1}}
\and S.\ Zucker\inst{\ref{inst2},\ref{inst1}} 
\and R.\ Helled\inst{\ref{inst3}}
\and D.\ Bashi\inst{\ref{inst4}}}

\institute{
School of Physics and Astronomy, Raymond and Beverly Sackler Faculty of Exact Sciences, 
Tel Aviv University, Tel Aviv, 6997801, Israel
\email{noybenyehuda@mail.tau.ac.il,shayz@tauex.tau.ac.il}
\label{inst1}
\and
Department of Geophysics, Raymond and Beverly Sackler Faculty of Exact Sciences, 
Tel Aviv University, Tel Aviv, 6997801, Israel
\label{inst2}
\and
Institut für Astrophysik, Universität Zürich, Winterthurerstr. 190, CH8057 Zurich, Switzerland
\email{ravit.helled@uzh.ch}
\label{inst3}
\and
Astrophysics Group, Cavendish Laboratory, University of Cambridge, JJ Thomson Avenue, Cambridge CB3 0HE, UK
\email{db975@cam.ac.uk
\label{inst4}}}

\begin{document}

\abstract{
The well-known correlation between stellar metallicity and planet occurrence is strongest for giant planets, but weaker for smaller planets, suggesting that detailed elemental patterns beyond $[\element{Fe}/\element{H}]$ may be relevant. Using abundances from the fourth data release of the GALAH spectroscopic survey, we analyzed $104$ host stars with $141$ confirmed transiting planets. We divide planets at \mbox{$r_{\mathrm{p}} = 2.6\,R_\oplus$}, the theoretical threshold radius above which planets are unlikely to be pure-water worlds. We find that large-planet hosts are enriched by $\mathord\sim 0.2$\,dex in iron and show a possible excess of highly volatile elements (\element{C}, \element{N}, \element{O}), though these measurements are affected by observational limitations, whereas small-planet hosts exhibit an enhanced contribution of the classical rock-forming elements (\element{Mg}, \element{Si}, \element{Ca}, \element{Ti}) relative to iron, corresponding to a modest $[\mathrm{Rock}/\element{Fe}]$ offset of $0.06$\,dex, which is statistically significant, with a $p$~value of $10^{-4}$. These offsets remain significant for alternative radius cuts. A matched control sample of non-planet-host stars shows only weak and mostly statistically insignificant similar trends, confirming that the stronger chemical signatures are linked to the planetary characteristics. As our study relies on transiting planets, it mainly probes short-period systems ($P<100$\,days). These results refine the planet–metallicity relation, highlighting the role of the relative balance between iron, volatiles, and rock-forming elements in planet formation.
}

\maketitle

\nolinenumbers

\section{Introduction}
\label{sec:intro}

A key empirical result in exoplanetary science is the correlation between stellar metallicity and planet occurrence. Radial-velocity surveys have shown that gas-giant planets are several times more common around stars with \mbox{$[\element{Fe}/\element{H}]\!\gtrsim\!+0.3$\,dex} than around solar-metallicity stars \citep{Sanetal2004,FisVal2005,Johetal2010}. This metallicity-occurrence relation provided crucial support for the core accretion paradigm, in which the enhanced heavy-element  content of a protoplanetary disk accelerates the formation of massive planetary cores before disk gas dispersal. In contrast, disk instability scenarios predict only a weak metallicity dependence, since giant planets are assumed to form through gravitational collapse of gas rather than core buildup \citep{Boss1997,Boss2002}. Thus, the observed trend is widely regarded as one of the strongest observational pillars of the core accretion framework.

With the advent of \textit{Kepler} and \textit{TESS}, it became clear that this trend extends to smaller planets as well, although with a weaker slope. The frequency of $1-5\,R_\oplus$ planets still rises with host-star metallicity, albeit more moderately than for gas giants \citep{WanFis2015}. Large occurrence studies have also shown that compact, close-in planets are ubiquitous even around stars of near-solar metallicity, suggesting that the formation of small planets requires less extreme heavy-element  enrichment \citep{Mayetal2009,Bucetal2012}. At the same time, population studies have revealed that small close-in planets are more common around $\alpha$-rich thin-disk stars, emphasizing that detailed stellar chemistry, not only overall metallicity, modulates planet formation efficiency \citep{BasZuc2019,Basetal2020}. These findings point toward the importance of considering more complex chemical characteristics rather than treating metallicity as a single number.

This view can be linked to the condensation-temperature framework of protoplanetary disks. Major compounds of different elements condense at characteristic temperatures in the protoplanetary disk \citep{Lod2003}, and the resulting volatile and refractory inventories provide the raw materials for planetary interiors. Highly volatile species such as carbon, nitrogen, and oxygen, dominate the composition of ices and atmospheres, while refractories such as magnesium, silicone, and calcium, form the planetary silicate mantle. Among these, iron plays a distinctive role: it is the only major condensate to form in metallic form at nebular temperatures, and it dominates the core mass fraction of rocky planets, strongly influencing their density and ability to retain atmospheres. Iron is also the element for which stellar abundances are measured most precisely, making $[\element{Fe}/\element{H}]$ both a convenient tracer and a physically meaningful parameter in studies of planet occurrence.

Meanwhile, it turned out that the distribution of planet sizes showed a pronounced “radius valley” near \mbox{$r_p\simeq 1.8-2.0 \, R_\oplus$} \citep{Fuletal2017,Vanetal2018,Macetal2023}, usually interpreted as the outcome of atmospheric mass-loss processes, either driven externally by stellar irradiation \citep[photoevaporation;][]{OweWu2013,LopFor2013} or internally by the cooling luminosity of the planetary core \citep[core-powered mass loss;][]{Ginetal2016,Ginetal2018}. Recent simulations also suggest involvement of orbital evolution, dynamical instabilities and late giant impacts in forming and shaping the radius valley \citep[e.g.][]{Venetal2020,Buretal2024,ShiIzi2025}. Interior modeling suggests that a related composition-driven transition occurs near $\mathord\sim\,2.6\,R_\oplus$, separating predominantly rocky planets from volatile-rich ones \citep{Lozetal2018}. Thus, both evolutionary and primordial explanations have been put forward, and the true origin of the valley is likely a combination of these mechanisms. Importantly, the depth and morphology of the valley itself depend on stellar chemistry: for example, it is more pronounced at higher metallicity and $\alpha$-enhancement \citep{Cheetal2022}, suggesting that chemical composition not only seeds planet formation but also influences long-term planetary evolution.

Large homogeneous spectroscopic surveys now make it possible to test these hypotheses systematically. While large compilations such as the Hypatia Catalog \citep{Hinetal2014,HinUnt2018,Hinetal2019} have provided valuable insights into host-star chemistry, their heterogeneous literature origins limit the precision of comparative studies. The GALAH survey \citep{Budetal2025} now offers homogeneous abundances for up to $32$ elements across nearly a million stars, enabling sensitive differential tests of volatile-to-refractory hypotheses at scale. In this work, we cross-match the catalog of the fourth data release of GALAH (GALAH~DR4) with confirmed exoplanet hosts to investigate how stellar abundances correlate with the sizes of the planets they host. We find that large-planet hosts appear enriched in iron and possibly in highly volatile elements, though the latter may be affected by observational limitations (Sect.~\ref{sec:flags}). This iron-dominated refractory enrichment, together with a potential volatile enhancement, refines the classical metallicity paradigm and highlights the role of detailed stellar chemistry in shaping planetary sizes (and therefore the planetary type).

The paper is organized as follows. Section~\ref{sec:data} describes the data and methodology, including the construction of the planet sample, the stellar abundances from GALAH~DR4, the composite abundance indices, the weighting scheme, the statistical tests, and the control-sample comparison. Section~\ref{sec:results} presents the results of the element-by-element analysis, the composite abundance indices, and the control-sample results. Section~\ref{sec:disc} discusses the implications and limitations of these findings, and Sect.~\ref{sec:summary} summarizes our conclusions.

\section{Data and methods}
\label{sec:data}

\subsection{Sample}
\label{subsec:stars}

We cross-matched the stars included in the GALAH~DR4 database \citep{Budetal2025} with the NASA Exoplanet Archive table of confirmed planets \citep{Chretal2025} on April 10th 2025, yielding $N\textsubscript{host}=104$ main-sequence planet-hosting stars and $N\textsubscript{pl}=141$ confirmed planets. Since our analysis requires precise radii, we are restricted to transiting planets. Consequently, our sample is inherently biased toward short-period systems; in practice, nearly all of our planets have orbital periods shorter than $100$ days.

Following the compositional transition identified by \citet{Lozetal2018}, we divided the sample at \mbox{$r_{\mathrm{p}} = 2.6\,R_\oplus$}. We labeled planets with \mbox{$r_\mathrm{p} < 2.6\,R_\oplus$} as \emph{small}, totaling $78$ planets, whereas those with \mbox{$r_\mathrm{p} \ge 2.6\,R_\oplus$} as \emph{large}, comprising $63$ planets. 

\subsection{Abundances}
\label{subsec:abunds}

The GALAH survey provides high-resolution ($R\simeq28000$) optical spectroscopy for nearly a million stars, with DR4 delivering abundances for up to $32$ elements. Typical uncertainties for well-measured species are $\mathord\sim\,0.03-0.07$ dex.

The catalog reports $[\element{Fe}/\element{H}]$ directly, while all other abundances were expressed relative to iron, $[\mathrm{X}/\element{Fe}]$\footnote{Throughout we adopt the solar reference scale used internally by GALAH.}. We recover absolute abundances as: 
\begin{equation*}
[\mathrm{X}/\element{H}]=[\mathrm{X}/\element{Fe}]+[\element{Fe}/\element{H}] \ .    
\end{equation*}

The GALAH limited availability of some of the abundance indices constrained the power of the single-element approach. To maximize sample size while preserving physical meaning, after a preliminary element-by-element study (whose results will be presented in Sect.~\ref{subsec:ebe}), we regrouped elements by condensation temperature, adopting the grouping suggested by \citet{Lod2003}, and keeping only elements that would not reduce the sample size too much:

Following \citet{Lod2003}, we collapsed the measured elements into four
condensate groups: (i) \textit{highly volatiles} -- \element{C}, \element{N}, \element{O}; (ii) \textit{volatiles} -- \element{C}, \element{N}, \element{O}, \element{Na}, \element{K}, \element{Cu}, \element{Zn}; (iii) \textit{refractories} -- \element{Mg}, \element{Al}, \element{Si}, \element{Ca}, \element{Sc}, \element{Ti}, \element{V}, \element{Cr}, \element{Fe}, \element{Co}, \element{Ni}, \element{Ba}; and (iv) \textit{rock-forming} -- \element{Mg}, \element{Al}, \element{Si}, \element{Ca}, \element{Ti}. 

We note that these groupings are not mutually exclusive. In particular, the highly volatile elements (\element{C}, \element{N}, \element{O}) also enter the broader “volatile” category, and the classical rock-forming species (\element{Mg}, \element{Si}, \element{Ca}, \element{Ti}) are a subset of the wider refractory pool. This overlap reflects the hierarchical nature of condensation chemistry: broader groups trace the bulk volatile or refractory budget, while the nested subgroups isolate specific components (e.g.\ the rock-forming silicates). By retaining this overlap, we can test both the overall volatile-refractory balance and the distinct role of individual subgroups within it.

To ensure statistical robustness, we restricted the group-composite indices to include only elements measured for a sufficiently large fraction of stars, so that the host-star sample retained more than $80$ stars for each composite index.

We excluded iron from our group of rock-forming elements since iron warrants special attention. It is the only major condensate to appear in metallic form at nebular temperatures ($\mathord\sim\,1350$\,K), it dominates the core mass fraction that controls the mass-radius relation of rocky planets \citep[e.g.][]{Seaetal2007,Valetal2007,Zenetal2016}, and owing to its rich line spectrum in cool, near-solar-metallicity stars -- the regime mainly probed by GALAH -- $[\element{Fe}/\element{H}]$ remains the most precise and externally calibrated tracer of stellar metallicity and the planet-occurrence trend.

GALAH reports the elemental abundances as $[\mathrm{X}/\element{Fe}]$ (and $[\element{Fe}/\element{H}]$) on the usual logarithmic scale. For our purposes, we combined individual elements into mass-weighted composite indices, following the prescription detailed by \citet{Gon2009}:

\begin{enumerate}

\item Convert each element abundance to linear, solar-normalized abundance: 
\begin{equation*}
R_\mathrm{X}=10^{[{\mathrm{X}/\element{H}}]}
\end{equation*}

\item Apply atomic-mass weighting and restore the solar scale: 
\begin{equation*}
M_\mathrm{X}=A_\mathrm{X}R_\mathrm{X}10^{A(\mathrm{x})-12}
\end{equation*}
where $A_\mathrm{X}$ is the atomic mass and $A(\mathrm{X})$ is the solar photospheric abundance from the “Reference” column of Table~8 by \citet{Budetal2025}.

\item Compute the solar reference value:
\begin{equation*}
M_{\mathrm{X},\odot} = A_\mathrm{X} 10^{A(x)-12}
\end{equation*}

\item Form the group composite (referring to the groups listed above) and return to the standard logarithmic form:
\begin{equation*}
[\mathrm{Group}/\element{H}] = \log_{10} \bigl( \sum_\mathrm{X} M_\mathrm{X} \bigr) - \log_{10} \bigl( \sum_\mathrm{X} M_{\mathrm{X},\odot} \bigr)
\end{equation*}

\item Calculate the ratio index between two groups, normalized to the solar reference:
\begin{equation*}
    [\mathrm{Group1}/\mathrm{Group2}] =
\log_{10} \left( \frac{M_{\mathrm{Group1,star}}}{M_{\mathrm{Group2,star}}} \right)
- \log_{10} \left( \frac{M_{\mathrm{Group1},\odot}}{M_{\mathrm{Group2},\odot}} \right) . 
\end{equation*}

\end{enumerate}

This procedure yields composite indices such as $[\mathrm{Vol}/\element{H}]$, $[\mathrm{Ref}/\element{H}]$, $[\mathrm{Rock}/\element{H}]$, and their ratios $[\mathrm{Vol}/\mathrm{Ref}]$ and $[\mathrm{Rock}/\element{Fe}]$. These indices summarize the relative enrichment of different condensation-temperature groups while minimizing the impact of missing or uncertain individual element measurements.

\subsection{Quality flags}
\label{sec:flags}
To further ensure sample integrity, we inspected the element-specific quality flags provided by GALAH~DR4 in the catalog \texttt{galah\_dr4\_allstar\_240705} \citep{Budetal2025}. For each element, a corresponding flag, \texttt{flag\_X\_fe}, accompanies the reported abundance ratio $[\mathrm{X}/\element{Fe}]$ and encodes possible issues in its determination. Values of \texttt{flag\_X\_fe = 0} denote reliable measurements, while non-zero values identify potential problems such as poor spectral fits, low SNR, or incomplete wavelength coverage. In constructing our main sample, we adopted only abundances with \texttt{flag\_X\_fe = 0} for all elements except carbon and nitrogen.

For \element{C} and \element{N}, particular caution is required. As explained by \citet{Budetal2025}, 
the flags \texttt{flag\_X\_fe = 32} and \texttt{33} indicate abundance determinations affected by incomplete spectral coverage or possible instrumental artifacts in the relevant molecular regions. Since the subset of stars with \texttt{flag\_c\_fe = 0} and \texttt{flag\_n\_fe = 0} constitutes only about $1\%$ of the entire GALAH~DR4 sample, restricting to those values would render the carbon and nitrogen datasets statistically negligible. We therefore retained stars with \texttt{flag\_c\_fe = 32} or \texttt{33} and \texttt{flag\_n\_fe = 32} or \texttt{33}, while treating the resulting volatile indices with due caution in the interpretation (Sect.~\ref{sec:summary}).

\subsection{Statistical treatment of multiplicity}
\label{subsec:mult}

Our cross-match between the NASA Exoplanet Archive and GALAH~DR4 yields a total of $141$ planets orbiting $104$ host stars. Because planetary systems often contain multiple detected planets, a straightforward analysis at the planet level would overweight such systems relative to single-planet hosts. To avoid this bias, we adopted a weighting scheme in which each host star contributed a total weight of unity, regardless of the number of planets it hosts:

If host star $j$ has $m_j$ planets, then each of its planets \mbox{$k=1,...,m_j$} was assigned weight $w_{j,k}=1/m_j$\,. This scheme ensured that the weights of all planets belonging to a given host would sum to unity:
\begin{equation*}
\sum_{k=1}^{m_j} w_{j,k} = 1
\end{equation*}
This way, every host star contributes equally to the statistics, while still allowing us to make use of all confirmed planets in the system. We verified that the main chemical trends reported in this study are robust to alternative schemes (such as assigning equal weight to each planet, or restricting the sample to a single planet per host).

\subsection{Weighted $t$ test}
\label{subsec:t-test}

To quantify the chemical differences between stars hosting small and large planets, we estimated differences in means of elemental abundances and composite indices using weighted two-sample Welch's $t$~test \citep[e.g.][]{Rux2006}, which is robust to unequal sample sizes and variances. For an abundance index $X$, the weighted test statistic is defined by:
\begin{equation}
  \label{eq:tstat}
  t =  \frac{\bar{X}_\mathrm{Large}-\bar{X}_\mathrm{Small}}
       {\sqrt{s_\mathrm{Large}^2/N^\mathrm{eff}_\mathrm{Large} +
              s_\mathrm{Small}^2/N^\mathrm{eff}_\mathrm{Small} } }, 
\end{equation}
where the weighted mean for a subsample is: 
\begin{equation*}
  \bar{X} = \frac{\sum w_{j,k} X_j}{\sum w_{j,k}}\ ,
\end{equation*}
the weighted variance is
\begin{equation*}
  s^2 =
  \frac{\sum w_{j,k} \bigl(X_j-\bar{X}\bigr)^2}
       {\sum w_{j,k}} \ ,
\end{equation*}
and the effective sample size (considering dependencies) is defined as
\begin{equation*}
  N^\mathrm{eff} = 
  \frac{\bigl(\sum w_\mathrm{j,k}\bigr)^2}
       {\sum w^2_{j,k}} \ .
\end{equation*}

\subsection{$p$-value significance}
\label{subsec:sig}
To avoid assumptions about the distributions of our samples we decided to use a permutation test in order to assess the significance of our statistics \citep{Goo1994}. In order to account for planet multiplicity we adapted the classic permutation test and used a clustered permutation test:
\begin{enumerate}
\setlength\itemsep{3pt}
  \item Group planets by host star; randomly shuffle the list of host stars, so that planets orbiting a certain star will all remain affiliated to the same star, albeit probably different from the original star.
  \item Re-compute $t$ (Eq.~\ref{eq:tstat}) for each of $10^6$ random permutations.
  \item The two-sided $p$ value is then the fraction of permuted $|t|$ values exceeding the originally observed value $|t_\mathrm{obs}|$.
\end{enumerate}

\subsection{Control sample}
\label{subsec:control}
To test whether the chemical differences we identify are indeed related to the presence of planets in different sizes or simply reflect biases caused by broader Galactic abundance trends, we constructed a control sample of stars without any detected planets (either transit or radial-velocity planets). A naive control sample drawn randomly from the field-star population would risk introducing systematic differences in stellar parameters. We therefore matched each host star to a control star from GALAH~DR4 catalog selected to be as similar as possible in key stellar properties.

The matching was performed in the parameter space of effective temperature ($T_\mathrm{eff}$), surface gravity ($\log g$), the metallicity ($[\element{Fe}/\element{H}]$), and apparent magnitude ($m_G$, taken as the \textit{Gaia}~DR3 G-band mean magnitude, provided in GALAH~DR4 as \texttt{phot\_g\_mean\_mag}). For each potential pair of host and control stars, we calculated a normalized Euclidean distance: 
\begin{multline*}
d^2 = 
\left( \frac{T_\mathrm{eff, host} - T_\mathrm{eff,control}}{\sigma_{T_\mathrm{eff}}} \right)^2 \\
+ \left( \frac{\log g_\mathrm{host} - \log g_\mathrm{control}}{\sigma_{\log g}} \right)^2
+ \left( \frac{m_{G,\mathrm{host}} -  m_{G,\mathrm{control}}}{\sigma_{m_G}} \right)^2 \ ,
\end{multline*}
where $\sigma_{T_\mathrm{eff}}$, $\sigma_{\log g}$, and $\sigma_{m_G}$ are the standard deviations of these parameters across the entire catalog. This normalization ensures that differences in scale between parameters are accounted for, and that each contributes equally to the distance metric. For each host star, the control star with the smallest normalized distance was selected to the control sample.

Each control star was used only once (matching without replacement), to ensure statistical independence and prevent oversampling of a small subset of field stars. Note again that in systems with multiple planets, the same control star was assigned to all occurrences of the host, preserving the weighting scheme described above.

\begin{figure*}[t]
  \sidecaption
  \raisebox{0.35cm}{
   \begin{minipage}[t]{0.68\textwidth} 
    \centering
    \begin{subfigure}[c]{0.49\textwidth}
      \includegraphics[width=\linewidth]{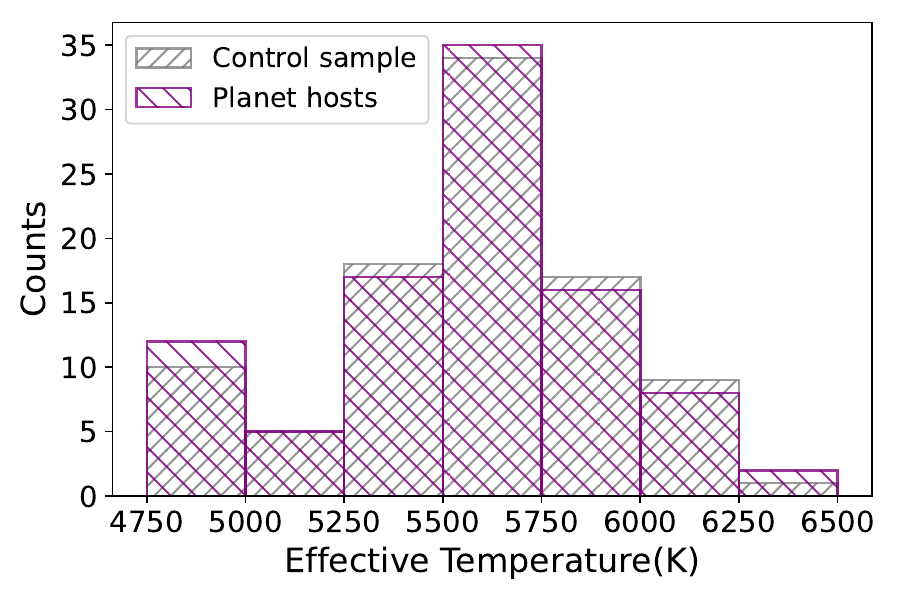}
    \end{subfigure}
    \begin{subfigure}[c]{0.49\textwidth}
      \includegraphics[width=\linewidth]{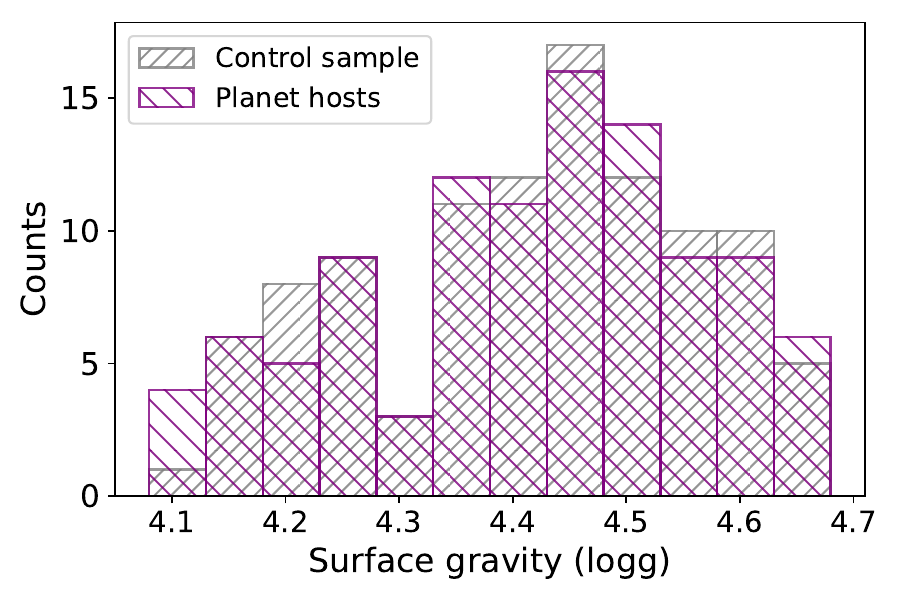}
    \end{subfigure}
    \vspace{1em}
    \begin{subfigure}[c]{0.49\textwidth}
      \includegraphics[width=\linewidth]{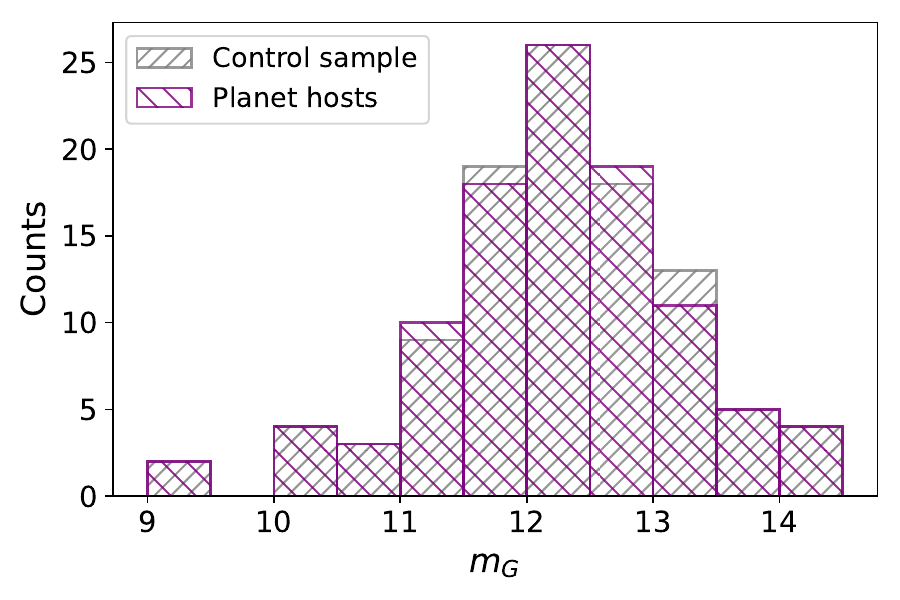}
    \end{subfigure}
    \begin{subfigure}[c]{0.49\textwidth}
      \includegraphics[width=\linewidth]{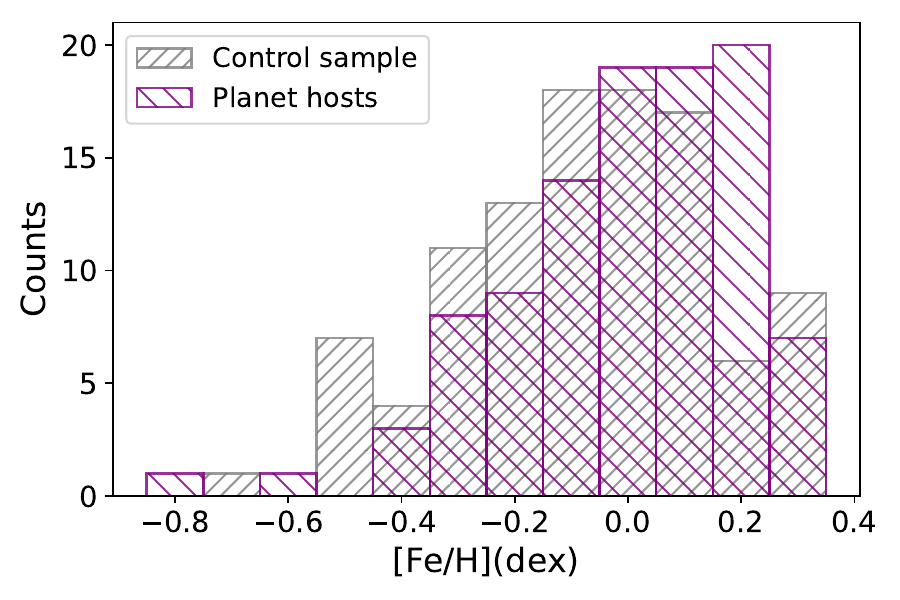}
    \end{subfigure}
  \end{minipage}
  }
  \caption{Validation of the control-sample matching. Distributions of effective
    temperature (upper left), surface gravity (upper right), \textit{Gaia} G-band
    apparent magnitude (lower left), and iron abundance (lower right) for the
    sample of planet-host stars and the matched control sample.}
  \label{fig:hostvcontrol}
\end{figure*}

After matching, we verified that the distributions of $T_\mathrm{eff}$, $\log g$ and $m_G$ in the control sample closely followed those of the host stars (Fig.~\ref{fig:hostvcontrol}). Weighted two-sample $t$~tests yielded $p$~values of $\sim 0.8-0.9$ in all three cases, indicating no statistically significant differences in means between the original planet-host sample and the matching control sample, in terms of these quantities. This agreement confirms that the procedure isolates abundance differences related to planet hosting, minimizing systematic biases caused by unrelated stellar parameter biases.

Furthermore, a potential bias could have arisen if a significant fraction of stars in either the test sample or the control sample were extremely iron-poor ($[\element{Fe}/\element{H} \lesssim -0.6$). Such stars are typically $\alpha$-enhanced thick-disk objects \citep[e.g.,][]{Edvetal1993,Benetal2003,Adietal2012}. As can be seen in Fig.~\ref{fig:hostvcontrol}, both our test and control samples contain very few such stars, and thus the statistical comparisons should not be significantly affected by this population effect.

\section{Results}
\label{sec:results}

\subsection{Stellar properties of subsamples}
\begin{figure*}[!htbp]
\centering
\begin{subfigure}[c]{0.33\textwidth}
    \includegraphics[width=\linewidth]{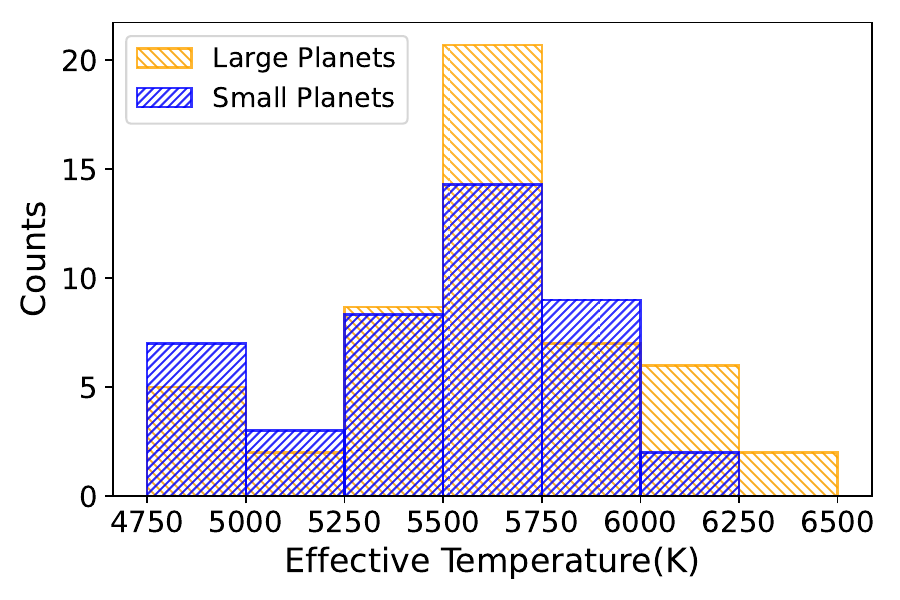}
\end{subfigure}
\hfill
\begin{subfigure}[c]{0.33\textwidth}
    \includegraphics[width=\linewidth]{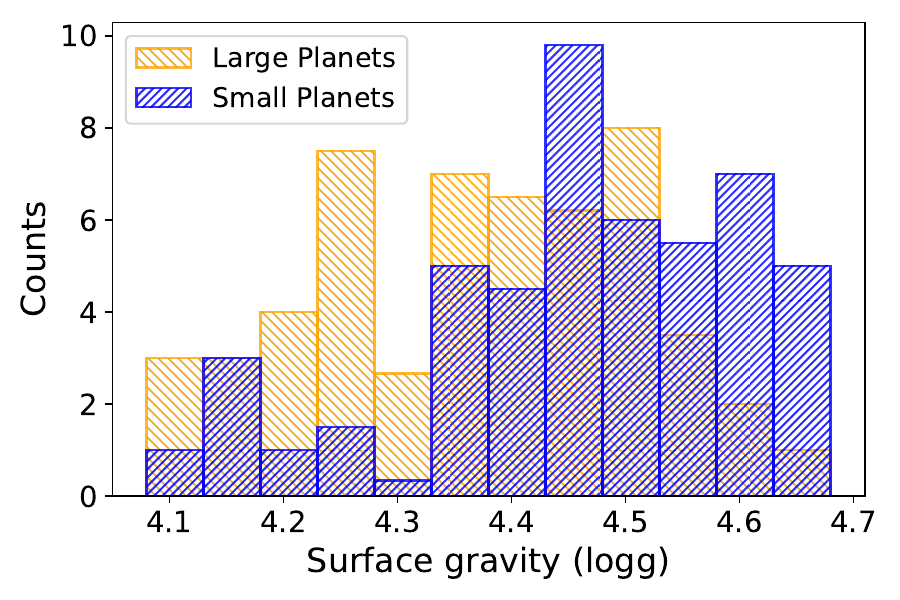}
\end{subfigure}
\hfill
\begin{subfigure}[c]{0.33\textwidth}
    \includegraphics[width=\linewidth]{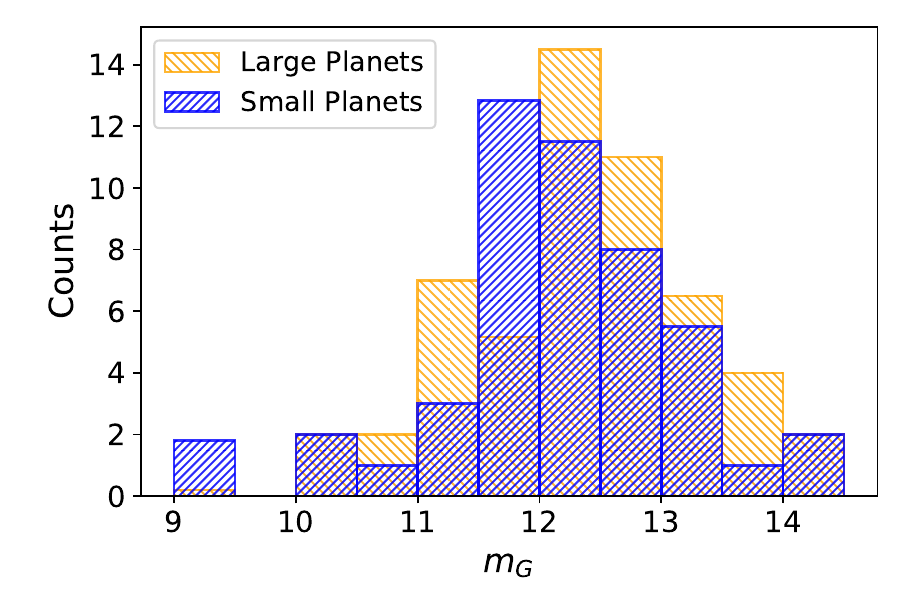}
\end{subfigure}
\hfill
\caption{Distributions of effective temperature (left), surface gravity (middle), and \textit{Gaia} G-band magnitude (right) for stars hosting large (orange) and small (blue) planets. Large-planet hosts are slightly hotter and have lower surface gravities on average, while the apparent-magnitude distributions are similar.}
\label{fig:subsampleshist}
\end{figure*}

Before comparing abundances, we assess whether stars hosting small and large planets occupy similar regions of stellar-parameter space. Figure~\ref{fig:subsampleshist} shows the distributions of $T_\mathrm{eff}$, $\log g$ and $m_G$ for the two subsamples. Weighted Welch $t$ tests of equality of means yield $p$ values of $0.006$ for $\log g$ (mean lower by $0.10$ for large-planet hosts) and $0.013$ for $T_\mathrm{eff}$ (mean higher by $224$\,K for large-planet hosts). The \textit{Gaia} magnitude $m_G$ does not exhibit a significant difference -- $0.11$\,mag, with a $p$~value of $0.356$. The modest but statistically significant offsets in effective temperature and surface gravity plausibly reflect detection-completeness effects -- transits of small planets are naturally shallower around larger, hotter stars.

\subsection{Element-by-element offsets}  
\label{subsec:ebe}

Table~\ref{tab:indiv_el} lists Welch's $t$-test $p$~values and mean differences between the small-planet and large-planet subsamples for $29$ individual abundance ratios (which were available for more than three host stars in our sample). The $p$~values were computed using the clustered permutation test on the weighted two-sample $t$~test as described above. The table also shows the differences in mean abundance between large-planet and small-planet hosts (computed as large minus small), and the number of host stars for which GALAH~DR4 contained an estimated abundance for the specific element. 

\begin{table}[!htbp]
\centering
\caption{Abundance offsets of small- and large- planet hosts.}
\begin{tabular}{|c|c|c|c|}
\hline
Element   & $t$-test   & $\Delta$Mean & Host star \\
abundance & $p$~value  &      (dex)   & sample size   \\
\hline
$[\element{C}/\element{Fe}]$\tablefootmark{$\dagger$} & 0.594046 & 0.02 & 104  \\
$[\element{N}/\element{Fe}]$\tablefootmark{$\dagger$} & 0.001409 & 0.26 & 104  \\
$[\element{O}/\element{Fe}]$ & 0.059510 & 0.11 & 104  \\
$[\element{Na}/\element{Fe}]$ & 0.000973 & 0.06 & 104  \\
$[\element{Mg}/\element{Fe}]$ & 0.000051 & -0.07 & 104  \\
$[\element{Al}/\element{Fe}]$ & 0.184435 & 0.03 & 97  \\
$[\element{Si}/\element{Fe}]$ &0.002423 & -0.04 & 99  \\
$[\element{K}/\element{Fe}]$ & 0.893489 & -0.01 & 104 \\
$[\element{Ca}/\element{Fe}]$ & 0.013277 & -0.04 & 104  \\
$[\element{Sc}/\element{Fe}]$ & 0.049160 & -0.03 &104\\
$[\element{Ti}/\element{Fe}]$ & 0.009402 & -0.05 & 90  \\
$[\element{V}/\element{Fe}]$ & 0.355588 & -0.03 & 97  \\
$[\element{Cr}/\element{Fe}]$ & 0.619519 & -0.01 & 104  \\
$[\element{Mn}/\element{Fe}]$ & 0.166730 & 0.03 & 104  \\
$[\element{Fe}/\element{H}]$ & 0.000016 & 0.15 & 104  \\
$[\element{Co}/\element{Fe}]$ & 0.049424 & -0.03 & 98  \\
$[\element{Ni}/\element{Fe}]$ & 0.010570 & 0.03 & 104  \\
$[\element{Cu}/\element{Fe}]$ & 0.047317 & 0.04 & 103  \\
$[\element{Zn}/\element{Fe}]$ & 0.254753 & -0.02 & 99 \\
$[\element{Rb}/\element{Fe}]$ & 0.054183 & -0.04 & 17  \\
$[\element{Sr}/\element{Fe}]$ & 0.517829 & -0.10 &5 \\
$[\element{Y}/\element{Fe}]$ & 0.835709 & -0.02 &83 \\
$[\element{Zr}/\element{Fe}]$ & 0.186629 & -0.10 & 22 \\
$[\element{Ru}/\element{Fe}]$ & 0.315930 & -0.20 & 4 \\
$[\element{Ba}/\element{Fe}]$ & 0.264789 & -0.03 & 104 \\
$[\element{La}/\element{Fe}]$ & 0.048538 & 0.11 & 16  \\
$[\element{Ce}/\element{Fe}]$ & 0.162651 & -0.11 & 9 \\
$[\element{Nd}/\element{Fe}]$ & 0.063066 & 0.07 & 41  \\
$[\element{Sm}/\element{Fe}]$ & 0.080839 & 0.20 & 7  \\
\hline
\end{tabular}
\tablefoot{The elements are ordered by increasing atomic number. \tablefoottext{$\dagger$}{Abundances of \element{C} and \element{N} are affected by lower reliability owing to flagged measurements in GALAH\,DR4 (see Sect.~\ref{sec:flags}). The corresponding results should therefore be interpreted with caution.}}
\label{tab:indiv_el}
\end{table}

Applying the False-Discovery Rate procedure \citep{BenHoc1995} for multiple comparisons, at a false-discovery rate of \mbox{$\alpha = 0.05$} (for a total of $29$ comparisons), leaves nine significant abundances:

\begin{itemize}
\setlength\itemsep{3pt}

  \item $[\element{N}/\element{Fe}]$ ($\Delta=+0.26$\,dex), the strongest volatile tracer;

  \item $[\element{Fe}/\element{H}]$ ($\Delta=+0.15$\,dex), which anchors the classic metallicity trend;
  
  \item $[\element{Na}/\element{Fe}]$ ($\Delta = 0.06$\,dex) -- a moderately volatile species;

  \item $[\element{Mg}/\element{Fe}]$, $[\element{Si}/\element{Fe}]$, 
  $[\element{Sc}/\element{Fe}]$, $[\element{Ca}/\element{Fe}]$, $[\element{Ti}/\element{Fe}]$, and $[\element{Ni}/\element{Fe}]$ ($\Delta \sim -0.07-0.03$\,dex) -- a generally coherent enhancement of the classical rock-forming and iron-peak inventory.

\end{itemize}
These nine element offsets point to a possible chemical dichotomy: large-planet hosts are enriched in iron and may show elevated volatile abundances, although the latter should be interpreted with caution (Sect.~\ref{sec:flags}). Small-planet hosts appear to be less iron-dominated, with higher ratios of rock-forming elements to iron. This points to a natural connection between enhanced rock-forming inventories and the assembly of smaller, predominantly rocky planets.

\subsection{Composite abundance ratios}
\label{subsec:composite}

\begin{figure*}[!htbp]
\centering

\vspace{2em}

\begin{subfigure}[c]{0.33\textwidth}
    \includegraphics[width=\linewidth]{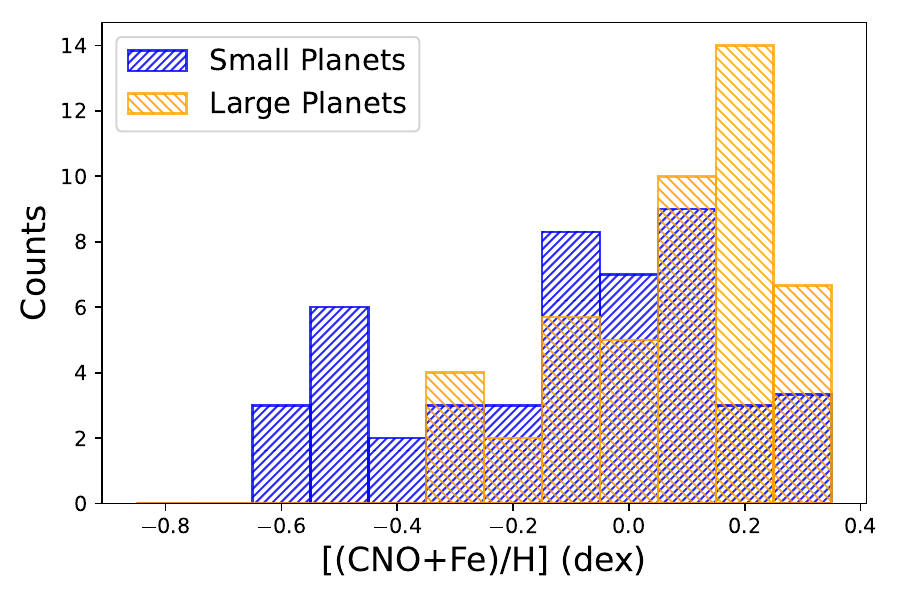}
\end{subfigure}
\hfill
\begin{subfigure}[c]{0.33\textwidth}
    \includegraphics[width=\linewidth]{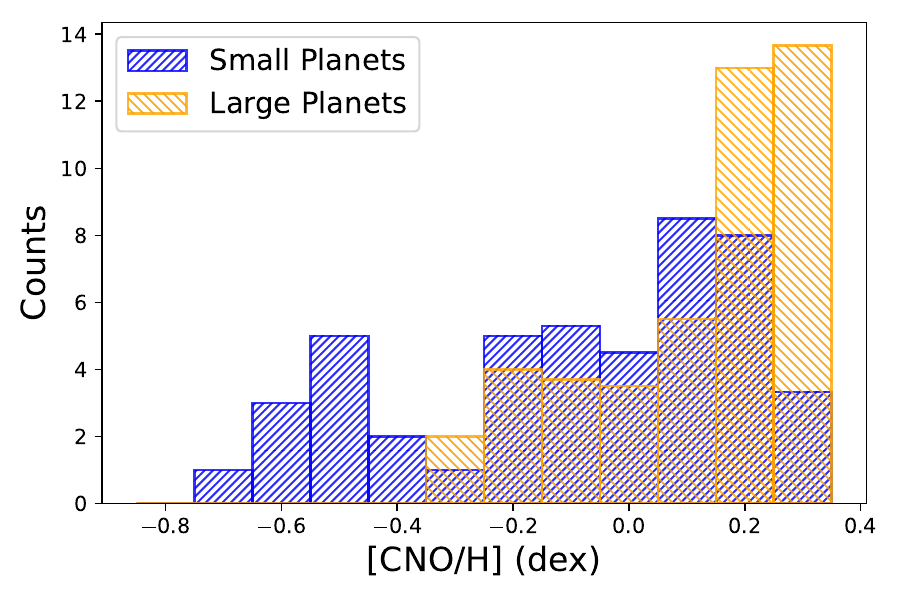}
\end{subfigure}
\hfill
\begin{subfigure}[c]{0.33\textwidth}
    \includegraphics[width=\linewidth]{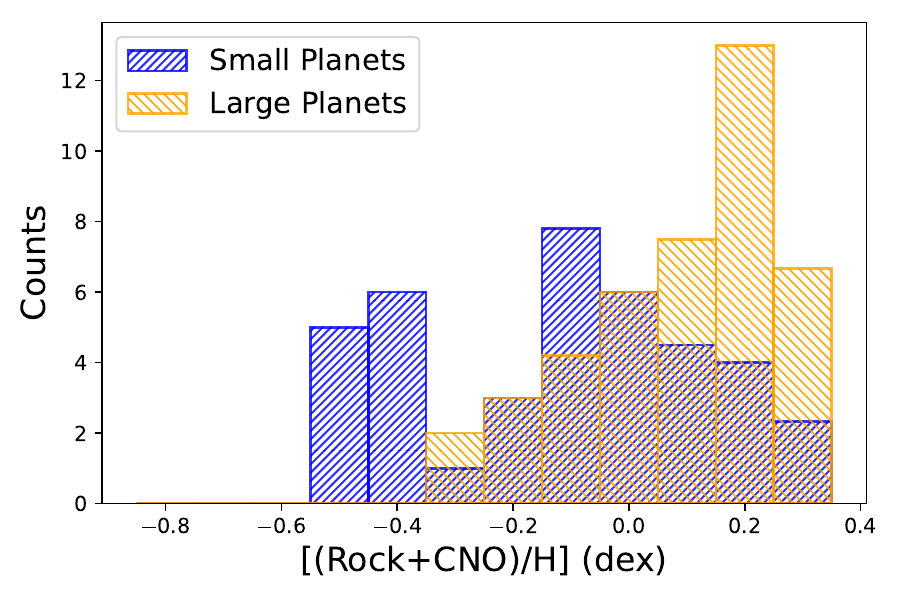}
\end{subfigure}
\hfill

\vspace{1em}

\begin{subfigure}[c]{0.33\textwidth}
    \includegraphics[width=\linewidth]{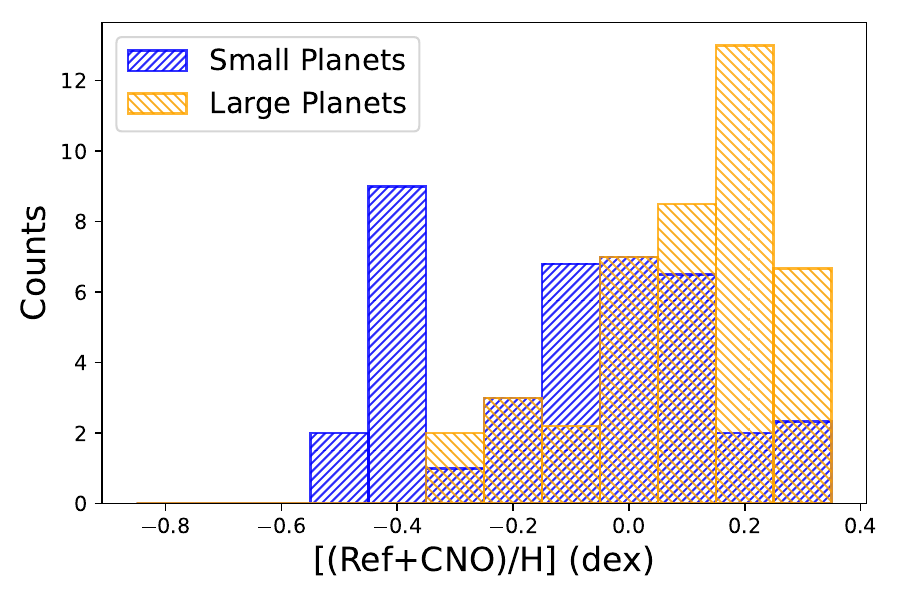}
\end{subfigure}
\hfill
\begin{subfigure}[c]{0.33\textwidth}
    \includegraphics[width=\linewidth]{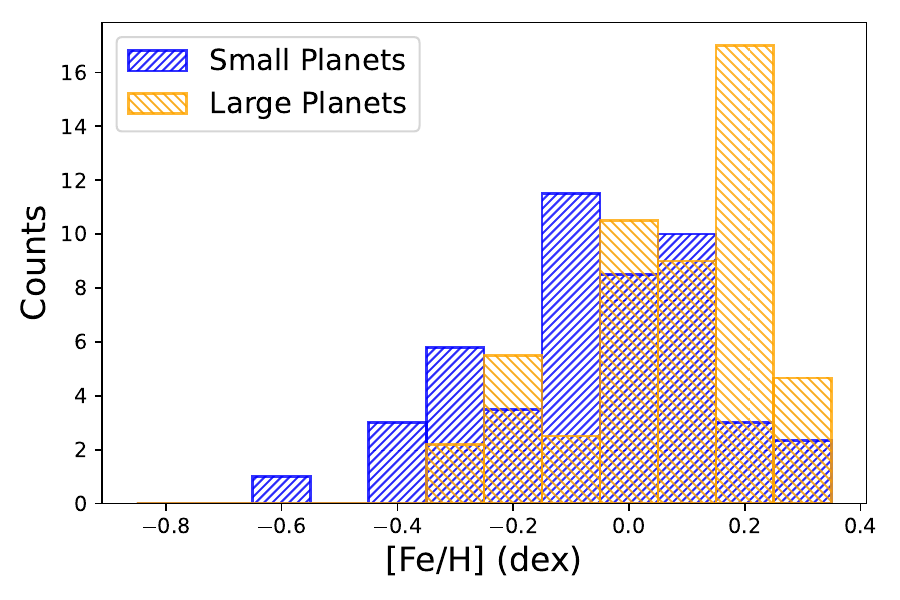}
\end{subfigure}
\hfill
\begin{subfigure}[c]{0.33\textwidth}
    \includegraphics[width=\linewidth]{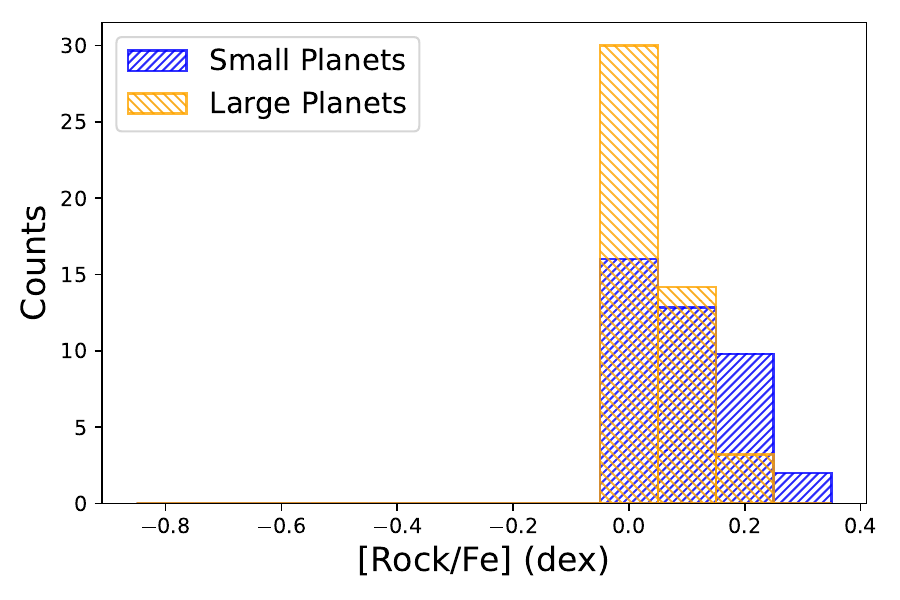}
\end{subfigure}
\hfill

\vspace{1em}

\begin{subfigure}[c]{0.33\textwidth}
    \includegraphics[width=\linewidth]{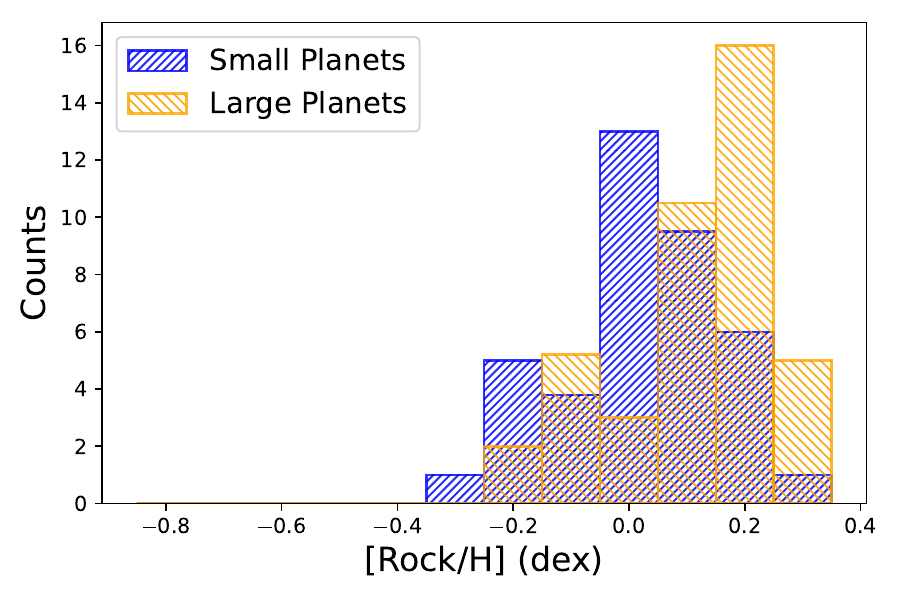}
\end{subfigure}
\begin{subfigure}[c]{0.33\textwidth}
    \includegraphics[width=\linewidth]{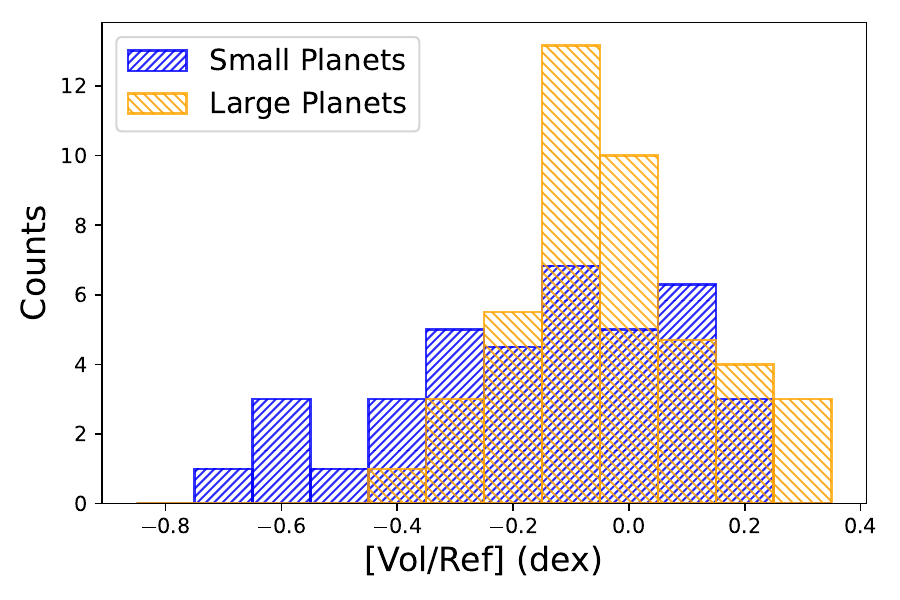}
\end{subfigure}
\hfill
\caption{Weighted histograms of the eight abundance indices in Table~\ref{tab:t-test_groups} for hosts of small (blue) and large (orange) planets; each star contributes unit total weight. CNO-based indices are affected by abundance quality flags (Sect.~\ref{sec:flags}) and the corresponding trends should therefore be interpreted cautiously.}
\label{fig:histograms}
\end{figure*}

Table~\ref{tab:t-test_groups} and Fig.~\ref{fig:histograms} compare eight abundance indices -- $[\element{Fe}/\element{H}]$ together with seven composite ratio indices of groups defined in Sect.~\ref{subsec:abunds} -- between the small-planet and large-planet subsamples. The histograms were constructed with the multiplicity-aware weighting described in Sect. \ref{subsec:mult}, so that each host star contributes equally.

The offsets and the $p$~values in Table~\ref{tab:t-test_groups} were estimated using the weighted Welch's $t$~test and the clustered permutation test as described in Sect.~\ref{sec:data}. The table also shows the sizes of the two subsamples for each element grouping, which depends on the availability of data pertaining to the comprising elements in GALAH~DR4. These quantities, $N_\mathrm{S}$ and $N_\mathrm{L}$, are actually the sums of weights in the small-planet and large-planet subsamples, corresponding to the total area under the histograms in Fig.~\ref{fig:histograms}. They are reported here as an intuitive measure of the effective sample coverage, and should not be confused with the formal effective sample sizes used in the $t$-test definition in Sect.~\ref{subsec:t-test}.

Most of the composites show highly statistically significant mean offsets of \mbox{$\mathord\sim\,0.20$\,dex} in the sense that large-planet hosts are more enriched. Two indices depart from this pattern.  $[\mathrm{Rock}/\element{H}]$ rises by only \mbox{$0.12$\,dex}, and thus $[\mathrm{Rock}/\element{Fe}]$ is in fact lower for large-planet hosts by \mbox{$0.06$\,dex}, revealing an iron-skew within the refractory pool. Finally, $[\mathrm{Vol}/\mathrm{Ref}]$ increases by \mbox{$0.13$\,dex}, and while it is still quite significant, with a $p$~value of $\mathord\sim\,0.003$, it is less significant than the other indices, perhaps due to a somewhat smaller sample of small-planet hosts.

\begin{table}[!htbp]
\centering
\caption{Composite abundance indices offsets between hosts of small and large planets. } 
\begin{tabular}{|l|c|c|c|c|}
\hline
Composite index & $t$-test  & $\Delta$Mean & $N_\mathrm{S}$ & $N_\mathrm{L}$ \\
                & $p$~value & (dex)        &                 & \\
\hline
$[(\mathrm{CNO}+\element{Fe})/\element{H}]$\tablefootmark{$\dagger$}  & 0.000005 & 0.23  & 49.63 & 54.36 \\
$[\mathrm{CNO}/\element{H}]$\tablefootmark{$\dagger$}                 & 0.000008 & 0.24  & 49.63 & 54.36 \\
$[(\mathrm{Rock}+\mathrm{CNO})/\element{H}]$\tablefootmark{$\dagger$} & 0.000009 & 0.23  & 41.63 & 47.36 \\
$[(\mathrm{Ref}+\mathrm{CNO})/\element{H}]$\tablefootmark{$\dagger$}  & 0.000009 & 0.22  & 40.63 & 47.36 \\
$[\element{Fe}/\element{H}]$                                          & 0.000016 & 0.15  & 49.63 & 54.36  \\
$[\mathrm{Rock}/\element{H}]$                                         & 0.000131 & 0.12  & 41.63 & 47.36 \\
$[\mathrm{Rock}/\element{Fe}]$                                        & 0.000138 & -0.06 & 41.63 & 47.36 \\
$[\mathrm{Vol}/\mathrm{Ref}]$\tablefootmark{$\dagger$}                & 0.003257 & 0.13  & 39.63 & 47.36 \\
\hline
\end{tabular}
\tablefoot{The number of host stars in each group is denoted as $N_\mathrm{S}$ for the small-planet sample and $N_\mathrm{L}$ for the large-planet sample. \tablefoottext{$\dagger$}{Indices involving \element{C} and \element{N} abundances are affected by lower reliability owing to flagged measurements in GALAH~DR4 (see Sect.~\ref{sec:flags}). The corresponding results should be interpreted with caution.}}
\label{tab:t-test_groups}
\end{table}

Since oxygen abundances are not affected by the \element{C} and \element{N} flags, we also examined oxygen-only composites. These show offsets of $\sim 0.20-0.22$\,dex with very low $p$~values, indicating that the volatile-like trend is not solely driven by the flagged \element{C} and \element{N} measurements.

\subsection{Control sample comparison}
\label{subsec:control_results}

To assess whether the abundance offsets identified above are intrinsic to planet hosts or somehow related to broader Galactic chemical trends, we repeated the analysis on the matched control sample described in Sect.~\ref{subsec:control}. In this case, each planet host was paired with a GALAH field star of similar $T_\mathrm{eff}$, $\log g$ and $m_G$, and the resulting subsamples were compared using the same weighted Welch’s $t$-test procedure.

\begin{table}[!htbp]
\centering
\caption{Composite abundance indices offsets for the matched control sample.}
\begin{tabular}{|l|c|c|c|c|c|}
\hline
Composite index & $t$-test  & $\Delta$Mean \\
                & $p$~value & (dex)        \\
\hline
$[{\rm (CNO+Fe)/H}]$\tablefootmark{$\dagger$}   & 0.098776 & 0.07  \\
$[{\rm CNO/H}]$\tablefootmark{$\dagger$}        & 0.069974 & 0.09  \\
$[{\rm (Rock+CNO)/H}]$\tablefootmark{$\dagger$} & 0.106421 & 0.07  \\
$[{\rm (Ref+CNO)/H}]$\tablefootmark{$\dagger$}  & 0.143724 & 0.06  \\
$[{\rm Fe/H}]$                                  & 0.821116 & -0.01 \\
$[{\rm Rock/H}]$                                & 0.586002 & -0.02 \\
$[{\rm Rock/Fe}]$                               & 0.208325 & -0.03 \\
$[{\rm Vol/Ref}]$\tablefootmark{$\dagger$}      & 0.018026 & 0.11  \\
\hline
\end{tabular}
\tablefoot{$^{\dagger}$\,Indices involving \element{C} and \element{N} abundances are affected by lower reliability owing to flagged measurements in GALAH~DR4 (see Sect.~\ref{sec:flags}).}
\label{tab:t-test_control}
\end{table}

Table~\ref{tab:t-test_control} summarizes the results for the composite indices of the control sample. In contrast to the clear and highly significant offsets found for the planet-host sample, the control stars show only small abundance differences, typically in the range of $0.06-0.11$\,dex. Most of these differences are found to be statistically insignificant. for example, $[\element{Fe}/\element{H}]$ differs by only $-0.01$\,dex between the “small” and “large” control subsamples, and the offsets in \mbox{$[(\mathrm{CNO}+\element{Fe})/\element{H}]$}, $[\mathrm{CNO}/\element{H}]$, and $[(\mathrm{Ref}+\mathrm{CNO})/\element{H}]$ are all below $0.1$\,dex, with \mbox{$p\gtrsim0.07$}. The only index with marginal significance is $[\mathrm{Vol}/\mathrm{Ref}]$, which increases by $0.11$\,dex ($p\simeq0.018$), though even here the effect is weaker than in the planet-host sample.

These results demonstrate that the strong abundance differences reported above are not generic features of the field-star population. While some residual offsets do appear in the control sample, they are smaller in amplitude and generally fail to reach statistical significance. This supports the conclusion that the more pronounced iron enrichments and potentially also volatiles, together with the iron-skew signature among refractories, are genuinely linked to the presence of planets and their sizes.

\subsection{Robustness to radius cutoff}

\begin{figure}[!htbp]
\centering    
\hspace*{-0.3cm}
\includegraphics[width=1.04\linewidth]{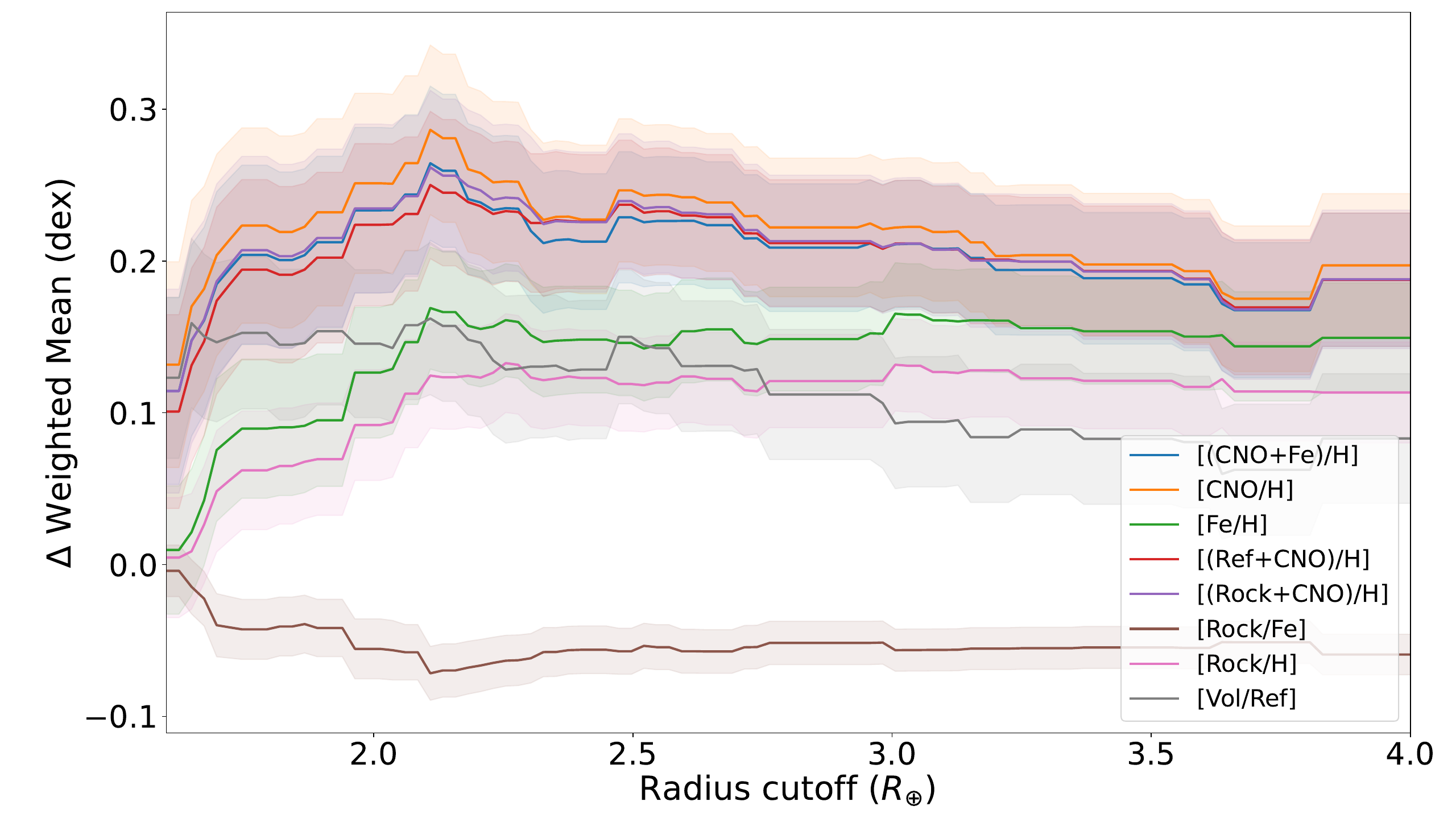}
\caption{Weighted mean abundance differences (large–small planet hosts) as a function of the chosen radius cutoff. Colored curves trace the various composite indices, with shaded bands showing uncertainties. The apparent volatile enhancement should be regarded as tentative given the limited reliability of \element{C} and \element{N} measurements.}
\label{fig:delta mean}
\end{figure}

Although we adopt $2.6 \, R_\oplus$ as a reference value, motivated by interior-structure models \citep{Lozetal2018}, our findings are not tied to this exact value. Figure~\ref{fig:delta mean} shows the mean abundance offsets between large-planet and small-planet hosts as a function of the cutoff radius. The main trends -- enrichment in iron and volatiles among large-planet hosts, and relatively stronger rock-forming inventories among small-planet hosts -- persist across cutoff values between $\sim2.0$ and $3.0 \, R_\oplus$. We note that we plot the abundance offsets rather than the associated $p$~values, since the latter mostly track the changing subsample size, whereas the offsets more directly illustrate the underlying chemical trends. Thus, toward the edges of the $2.0 - 3.0 \, R_\oplus$ range the statistical significance weakens, but this is related to the smaller subsample size. Overall the chemical dichotomy remains. We return to the implications of this test in Sect.~\ref{sec:disc}.

\section{Discussion}
\label{sec:disc}

Our results extend the classical planet–metallicity correlation \citep{Sanetal2004,FisVal2005} in two important directions. First, they confirm iron enrichment as a hallmark of large-planet hosts, consistent with the idea that a high heavy-element budget facilitates the assembly of massive cores capable of retaining thick atmospheres. Second, we find tentative evidence for enrichment in the highly volatile elements carbon, nitrogen, and oxygen, although this result relies on flagged abundance measurements (Sect.~\ref{sec:flags}) and should be interpreted with caution. In contrast, stars hosting small planets display a stronger relative contribution of the classical rock-forming elements (magnesium, silicone, calcium, titanium).

This chemical dichotomy resonates with the morphology of the exoplanet radius distribution. The “radius valley” near $1.8 - 2.0 \, R_\oplus$ is usually attributed to atmospheric mass loss \citep[e.g.][]{OweWu2013,LopFor2013,Ginetal2016}. However, interior models suggest that a more fundamental composition-driven transition occurs near $\mathord\sim2.6 \, R_\oplus$ \citep{Lozetal2018}, separating rocky from volatile-rich planets. Our findings are consistent with this picture: systems enriched in iron and possibly volatiles preferentially host planets above the break, while those with enhanced rock-formers relative to iron are more likely to host planets below it. Stellar chemistry thus appears to shape not only the efficiency of planet formation, but also the eventual distribution of planetary sizes.

We tested the sensitivity of these trends to varying the planet–radius cutoff. While statistical significance naturally decreases as the sample sizes change, the qualitative separation between the two planet-host-stars subsamples persists, especially between $2.0$ and $3.0 \, R_\oplus$. However, because the signals remain similar regardless of the adopted boundary, the present sample is not rich enough to constrain the exact location of the cutoff radius itself. In this sense, our results point to the existence of a chemically defined transition, but do not constrain its exact location.

As a further robustness test, we compared our host sample with a matched control sample of non-host stars. The matched control sample of non-planet-hosting stars shows only weak and mostly insignificant versions of the same patterns, suggesting that the stronger signals in the host sample are genuinely linked to the presence and sizes of planets, rather than to Galactic chemical evolution.

Some limitations must be acknowledged. The host-star sample is modest ($104$ stars), which restricts the ability to resolve finer subdivisions (e.g.\ by orbital period or multiplicity). Second, while GALAH~DR4 provides abundances for up to $32$ elements, coverage is incomplete and uncertainties vary, leading to sample-size differences between indices. In addition, the individual \element{C} and \element{N} abundances in GALAH~DR4 carry specific quality flags (Sect.~\ref{sec:flags}), and hence, trends involving CNO should be regarded as tentative. A similar offset is obtained when using oxygen alone, whose abundances are unflagged, suggesting that the volatile-related signal is not exclusively dependent on the \element{C} and \element{N} determinations.

We also examined each abundance index as a function of $T_\mathrm{eff}$ and $\log g$. These checks revealed mild parameter-dependent trends that were present in both the host and control samples. Importantly, the chemical offsets between large-planet and small-planet hosts persist even after accounting for the small differences in $T_\mathrm{eff}$ and $\log g$, indicating that these stellar-parameter variations do not explain the observed abundance differences. In our preliminary explorations of GALAH~DR4 we also saw hints of a possible dependence of planet size on stellar age, although the substantial age uncertainties and biases prevented a meaningful analysis at this stage. This remains an interesting direction for future work.

Note that our study is limited to transiting planets, and therefore predominantly probes short-period systems ($P<100$\,days). Whether the same chemical trends extend to longer-period planets requires additional data and remains to be tested in future research. Furthermore, small-planet transits are shallower around larger, hotter stars, which reduces completeness -- the lower $\log g$ and higher $T_\mathrm{eff}$ for large-planet hosts are therefore consistent with selection effects rather than intrinsic evolutionary differences between the host populations.

\section{Summary and Conclusions}
\label{sec:summary}

We analyzed chemical abundances from GALAH~DR4 for $104$ host stars with $141$ confirmed transiting planets, divided into “small” and “large” planets using a working threshold of \mbox{$r_\mathrm{p} = 2.6 \, R_\oplus$}. Our main findings are:

\begin{enumerate}
\setlength\itemsep{3pt}

\item Stars hosting large planets tend to be enriched in iron and show tentative indications of higher abundances of the highly volatile elements (\element{C}, \element{N}, \element{O}), while small-planet hosts appear less iron-dominated, with higher ratios of rock-forming refractories (\element{Mg}, \element{Si}, \element{Ca}, \element{Ti}) to iron. 

\item The results remain statistically significant for nearby choices of the dividing radius, but the current sample is too small to determine the precise location of the chemical transition.

\item A matched control sample of field stars shows only weak and mostly insignificant abundance trends, confirming that the stronger offsets in the host sample are not caused by the underlying stellar-population trends. 
\end{enumerate}

Our results refine the classical planet-metallicity relation. While iron remains central to the planet-metallicity relation, its influence is intertwined with volatile and rock-forming abundances. This chemistry-driven perspective situates exoplanet demographics within the broader framework of Galactic stellar populations and provides a path toward predicting planetary architectures from stellar abundances. We can therefore conclude that the outcome of planet formation also depends on the balance between iron, volatiles, and rock-forming elements. 

Finally, we emphasize an important limitation: although our analysis clearly indicates iron enrichment, and tentatively a volatile enhancement, the CNO abundances in GALAH~DR4 carry significant quality flags (Sect.~\ref{sec:flags}) and thus remain less reliable. The volatile signal should therefore be regarded as provisional. However, oxygen-only indices, which are unaffected by the \element{C} and \element{N} observational issues, show comparable $\mathord\sim0.2$\,dex offsets and thus provide partial support for a volatile-related trend.

In contrast to volatiles, the $[\mathrm{Rock}/\element{Fe}]$ offset, quantifying the excess of classical rock-forming elements (\element{Mg}, \element{Si}, \element{Ca}, \element{Ti}) relative to iron by $\lesssim 0.1$\,dex, is statistically robust and unaffected by these caveats. This persistent ratio suggests that the balance between silicate-forming material and metallic iron, rather than the uncertain volatile content, provides the most reliable chemical fingerprint associated with planet size, and motivates future high-precision abundance analyses of CNO elements to test this emerging picture. 

Future surveys such as \textit{PLATO}, \textit{Roman}, and ELTs can test these trends using larger and more diverse stellar populations, and enhance our understanding of how stellar composition influences planet formation.  

\begin{acknowledgements} 
This research was supported by the ISRAEL SCIENCE FOUNDATION (grant No. 1404/22). This research has made use of the NASA Exoplanet Archive, which is operated by the California Institute of Technology, under contract with the National Aeronautics and Space Administration under the Exoplanet Exploration Program. This work made use of the Fourth Data Release of the GALAH Survey. The GALAH Survey is based on data acquired through the Australian Astronomical Observatory, under programs: A/2013B/13 (The GALAH pilot survey); A/2014A/25, A/2015A/19, A2017A/18 (The GALAH survey phase 1); A2018A/18 (Open clusters with HERMES); A2019A/1 (Hierarchical star formation in Ori OB1); A2019A/15, A/2020B/23, R/2022B/5, R/2023A/4, R2023B/5 (The GALAH survey phase 2); A/2015B/19, A/2016A/22, A/2016B/10, A/2017B/16, A/2018B/15 (The HERMES-TESS program); A/2015A/3, A/2015B/1, A/2015B/19, A/2016A/22, A/2016B/12, A/2017A/14, A/2020B/14 (The HERMES K2-follow-up program); R/2022B/02 and A/2023A/09 (Combining asteroseismology and spectroscopy in K2); A/2023A/8 (Resolving the chemical fingerprints of Milky Way mergers); and A/2023B/4 (s-process variations in southern globular clusters). We acknowledge the traditional owners of the land on which the AAT stands, the Gamilaraay people, and pay our respects to elders past and present. This paper includes data that has been provided by AAO Data Central (datacentral.org.au). We thank the anonymous referee for the wise and constructive comments, which improved this paper substantially.
\end{acknowledgements}

\bibliographystyle{aa}
\bibliography{volhosts}
 
\end{document}